\documentclass[12pt]{article}
\usepackage[dvips]{graphicx}

\usepackage{color}
\usepackage{lscape}
\usepackage[x11names,rgb]{xcolor}

\usepackage{tikz}
\usetikzlibrary{arrows,snakes,backgrounds}

\usepackage{amsfonts}
\usepackage{amsmath}
\usepackage{amssymb}
\usepackage{amsthm}
\usepackage{enumerate}

\usepackage[pdftex]{hyperref}
\usepackage{natbib}
\bibpunct{(}{)}{;}{a}{,}{,}
\hypersetup{colorlinks,%
            citecolor=blue,%
            filecolor=blue,%
            linkcolor=blue,%
            urlcolor=blue,%
            }

\renewcommand{\vec}[1]{\mathbf{#1}}

\theoremstyle{remark} \newtheorem{example}{{\bf \sc Example}}

\newcommand{\toprule}{\hline}
\newcommand{\midrule}{\hline}
\newcommand{\bottomrule}{\hline}

\begin{document}

\title{A network centrality method for the rating problem}

\newcounter{savecntr}
\newcounter{restorecntr}

\author{
Yongli Li$^{* , \dag}$
\and 
Paolo Pin\thanks{%
Dipartimento di Economia Politica e Statistica, Universit\`a degli Studi di Siena, Italy.
Email: paolo.pin@unisi.it.}
\and
Chong Wu\thanks{%
School of Management, Harbin Institute of Technology, Harbin 150001, P.R.China}}
\date{March 2014}

\maketitle

\begin{abstract}
We propose a new method for aggregating the information of multiple reviewers rating multiple products.
Our approach is based on the network relations induced between products by the rating activity of the reviewers.
We show that our method is algorithmically implementable even for large numbers of both products and consumers, as is the case for many online sites.
Moreover, comparing it with the simple average, which is mostly used in practice, and with other methods previously proposed in the literature, it performs very well under various dimension, proving itself to be an optimal trade--off between computational efficiency, accordance with the reviewers original orderings, and robustness with respect to the inclusion of systematically biased reports.
\end{abstract}

\bigskip


\newpage 

\section{Introduction}
\label{sec:intro}

When many reviewers rate goods or projects, the exercise of aggregating all this information is a useful one:~it helps consumers or principals to select the expected best projects, and induces an objective \emph{price} for each good that will smooth and make efficient any market procedure.
How to deal with this theoretical problem is an old issue in the literature on voting, stemming from \cite{arrow1963social}, and has been object of discussion for many real world applications, as the one of evaluating scientific research (\citealp{cook2005optimal}).
Nowadays, it has become a compelling exercise for big online sellers, when reporting huge feedback data from consumers.
Being able to offer a reliable aggregate ranking benefits the consumers to recognize favorite goods, and it is a service provided e.g.~by Amazon (\url{www.amazon.com}), Ebay (\url{www.ebay.com}) and Taobao (\url{www.taobao.com}).
Moreover, it is actually the core of the service of other online sites, as Tripadvisor (\url{www.tripadvisor.com}).
Many works in the recent past have demonstrated the significance of scores and remarks given by the online shoppers: some of them are \cite{ba2002evidence}, \cite{pavlou2004building}, \cite{pavlou2007understanding},  \cite{park2009effects}.

\bigskip

 Given a set of commodities, a set of customers and the rating scores of each customer-commodity pair, the general \emph{rating problem} is to rate each commodity with a single score. In this context, this paper aims to present a reasonable \emph{rating method}, implementable in efficient polynomial time by an algorithm, and whose results can be in accordance with most of customers' rankings.

\bigskip

The commonly adopted rating method in those real world applications is the averaging (maybe weighted -- we come back to weighted average in the conclusion), which is implementable in linear time, but has been found defective. In particular, the result of the averaging is likely to violate most customers' preferences (this is an issue already recognized by \citealp{arrow1963social}), even if a coherent ranking is actually available (as in Example \ref{example:no_coherent} of this paper).\footnote{%
See \cite{basili2013aggregation} for a recent analysis of this theoretical problem.}
On the other hand, more complicated methods must take into account other constraints.
One is that in many cases most of the reviewers rate only a small fraction of all the available goods, and it becomes useful to take into account also the weight of all this missing information.
Another one is that the computational complexity of every aggregate rating must be taken seriously into account, because the numbers of products and reviewers can be in the order of hundreds of thousands.
So, it is important to find a good trade off that is actually implementable by the online sites.

\bigskip

We put the focus of a new rating method, that we call the \emph{network centrality} method, on the linkages established by the customers: that is, the customers' rating actions make two commodities related the more consumers compare both of them. If the commodities are treated as nodes and the above linkages as edges, a weighted network will be obtained which can comprehensively reflect the aggregate information. 
We  do so borrowing from the literature on complex networks that have analyzed the importance of the \emph{centrality} of a node. 
We adopt the Bonacich--Katz centrality (see \citealp{bonacich2001eigenvector} for a fairly recent exposition) to define a \emph{network centrality} method of aggregate ranking.\footnote{%
See \cite{BKD_AER} for a recent discussion of the applications of the Bonacich--Katz centrality to economic environments with peer--effects.}

\bigskip

Our method is not the first one to address this issue.
The underlying idea of our method is based on the properties of a matrix representing the relations between products and reviewers.
The first method based on spectral analysis is the Analytical Hierarchy Process proposed by \cite{saaty1977scaling}, which requires that all the reviewer rate all the products.
More in general, methods based on eigenvector centrality (as the one used by \citealp{keener1993perron}), are unstable, as has been shown in the literature on peer effects in social networks (more on this in Section \ref{sec:beta}).

With a different approach,
\cite{kemeny1962preference} proposed an algorithm to minimize the aggregate discrepancy of an overall ranking with respect to each individual ranking, but their algorithm is NP-hard and then not feasible for instances with many products and reviewers.
\cite{hochbaum2006methodologies} and \cite{hochbaum2011rating} propose an approximation algorithm that works in polynomial time and approximates the one of \cite{kemeny1962preference}.
In this paper we actually  adopt the method of \cite{hochbaum2011rating} as a comparison with respect to our method, and the objective measure they minimize as one of the benchmarks of evaluation.
We also discuss why our method is not computationally worse, and is less demanding in terms of memory storage.

\bigskip

Another important issue to consider in the online rating applications is the following.
In some cases true or fake reviewers could be maliciously biased in favor of some specific products, and this is also an issue to consider.
This can happen because single fake accounts, known as \emph{Sybils} (see \citealp{wang2012social} and \citealp{cao2012aiding}), are created; or many of them are systematically included in the system to force an intentionally biased evaluation (a phenomenon called \emph{Crowdturfing}, see \citealp{wang2012serf}).
In such cases the simple rating method of averaging is considered a much better solution than the methods that preserve ordering, as averaging can be thought as an unbiased estimator of the \emph{true value} of the products when evaluations are affected by some \emph{white noise} (on this, see \citealp{girotra2010idea}).
However, the noise will have a well specified predetermined sign if it comes from intentional manipulation, and we show that in some cases our  method can overcome this unbiased malicious noise even better than the averaging.

\bigskip

Summing up, our network centrality method results to have very desirable features.
First,  it is algorithmically easy to compute, if compared to other measures in the literature.
Then, it performs very well in maintaining most of the original rankings of the reviewers, both on randomly generated synthetic data, and on real data from an online rating platform.
Third, it is robust to the artificial insertion of consumers systematically reporting fake data.

\bigskip

In Section \ref{sec:motivation} we provide a motivating intuition for our method, and describe it formally.
In  Section \ref{sec:simulations} we report results of an extensive numeric analysis that compares our method with others.
In Section \ref{sec:realdata} we  apply our method to a real online dataset where people rate movies.
Section \ref{sec:conclusion} concludes.

\section{Intuition and theory for our approach}
\label{sec:motivation}

From now on we call \emph{goods} the items to be evaluated, being projects or commodities, and \emph{agents} the reviewers that independently evaluate a subset of the goods. We then consider implicitly a principal that wants to aggregate all this information in a single vector assigning a score to each good. This will be the \emph{rating problem}.
A \emph{rating method} is an algorithm that provide a solution.
We will consider different \emph{measures} to evaluate rating methods.

\subsection{Intuition}
\label{sec:intuition}

\begin{figure}[h]
\begin{center}
 \includegraphics[width=10cm]{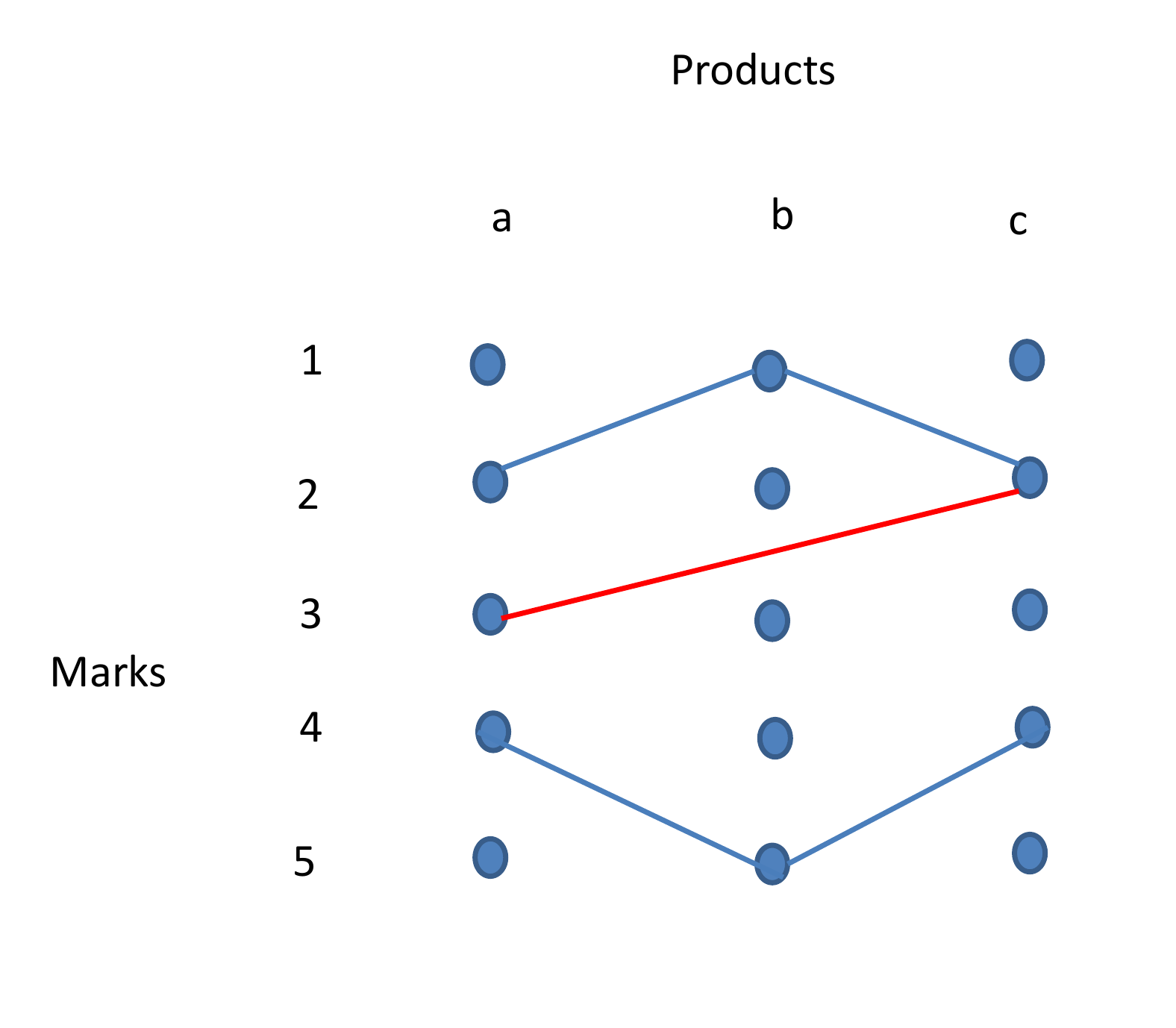}
\caption{The network representation of Table \ref{table:1} (the blue lines) and Table \ref{table:2} (including the red line).} \label{fig:1}
\end{center}
\end{figure}

Let us start with an example that provides the intuition for our method.
In a rating environment where agents assign marks (from 1 to 5) to products, consider the  situation illustrated in Table \ref{table:1}.
\begin{equation} \label{table:1}
\begin{array}{r|cccc|}
 & \multicolumn{4}{c}{Agents} \\
Products & 1 & 2 & 3 & 4 \\
\hline
a & 2 & - & 4 & - \\
b & 1 & 1 & 5 & 5 \\
c & - & 2 & - & 4 \\
\hline
\end{array} \ \ . 
\end{equation}
What is the score that we should attribute to product $b$ from this table?
The situation here is fully symmetric between two agents attributing a mark of $1$, and the other two attributing a mark of $5$.
This symmetry is not only in the grades they attribute to good $b$, but also in the overall marking above all products, as depicted by the blue lines in Figure \ref{fig:1}: this is a network where nodes are all possible marks for all goods, and there is a link between two nodes if at least one agent gave those specific two marks to those two goods.

\bigskip

Now suppose that a new agent 5 enters and assigns marks only to products $a$ and $c$, as depicted in Table \ref{table:2}, where a new column has been added.
\begin{equation} \label{table:2}
\begin{array}{r|ccccc|}
 & \multicolumn{5}{c}{Agents} \\
Products & 1 & 2 & 3 & 4 & 5\\
\hline
a & 2 & - & 4 & - & 3 \\
b & 1 & 1 & 5 & 5 & -\\
c & - & 2 & - & 4 & 2\\
\hline
\end{array} \ \ .
\end{equation}
Apparently this adds no information on the value of product $b$.
However, if we look at Figure \ref{fig:1} this breaks symmetry: now this new agent (the red link) agrees on product $c$ with an agent that gave a low mark to product $b$.
If we want to assign some value to this new piece of information we will give more weight to mark $1$ for good $b$, or equivalently a higher weight to node $b1$ in the network of Figure \ref{fig:1}.

\subsection{Formal Model}
\label{sec:model}

In general, a rating problem can be formalized and solved in the following way.
Consider $M$ goods that receives marks from $1$ to $a \in \mathbb{N}$, from $N$ agents.
An agent $n$ will typically assign marks only to a subset of $k_n$ out of the $M$ products, and consequently assign no mark (the $-$ symbol) to all the other $M-k_n$ products.
We will say that $xi \in S_n$ if agent $n$ assigns mark $i$ to product $x$.
This environment can be represented by an $N \times M$ matrix with elements from $\{1, \dots , a\} \cup \{ - \}$, which will typically be sparse (i.e.~with many `$-$' elements).
We call $\bf r$ this matrix, so that $r_{i,m}$ is the mark (or the $-$ symbol) that agent $i$ assigns to good $m$.\footnote{%
We will come back to this representation when analyzing other ranking measures in the literature, in Section \ref{sec:efficiency}.}
Another way to represent this environment is with an undirected weighted network with $a \cdot M$ nodes (one for every possible non--blank mark for each product) and where links are weighted in the following way.
The weight of a link between node $xi$ and different node $yj$ is given by the formula
\begin{equation}
\label{eq:link_value}
\ell_{xi,yj} \equiv \sum_{n \in N}  \frac{1}{k_n - 1} \left( \mathbb{I}_{xi \in S_n} \cdot \mathbb{I}_{yj \in S_n}  \right) \ \ ,
\end{equation}
where $\mathbb{I}_{x \in X}$ is the indicator function that has value $1$ if $x$ is an element of $X$, and has value $0$ otherwise.
We set by definition $\ell_{xi,xi}=0$, so that there are no links from any node to itself.
What formula (\ref{eq:link_value}) says is that if an agent $n$, who already marked $k_n$ products, assigns mark $i$ to good $x$, then this will add a value of $\frac{1}{k_n}$ to each link between node $xi$ and all those nodes $yj$ already assigned by agent $i$ and present in the set $S_n$.
Algebraically this just adds an aggregate value of $1$ to all the links of node $xi$.
We call $L$ the symmetric $aM \times aM$ \emph{adjacency matrix} obtained from (\ref{eq:link_value}).

If we sum any row or column of this matrix, say the one labelled $xi$, the result is
\[
\sum_{yj \ne xi} \sum_{n \in N}  \frac{1}{k_n - 1} \left( \mathbb{I}_{xi \in S_n} \cdot \mathbb{I}_{yj \in S_n} \right) = \sum_{n:~ xi \in S_n}  \frac{k_n - 1}{k_n - 1} = | \{ n:~ xi \in S_n \} | \ \ ,
\]
where the first passage is due to the fact that each agent rating $xi$ puts node $xi$ in relation with other $k_n - 1 $ nodes.
So, the sum on each row or column of  matrix $ L$ is just the number of agents that actually rated good $x$ with mark $i$.
In other words, in this network the \emph{degree} of a node is just the amount of $i$ marks provided to $x$ by the $N$ agents.
However, as is well known from network theory (see e.g.~\citealp{new1} or \citealp{Jackson_book}), the degree of a node is only one piece of information about its role in the network structure.
Going back to the example in Figure \ref{fig:1}, even if nodes $b1$ and $b5$ have the same degree, node $b1$ is more \emph{central} in the network, because there are more other nodes \emph{connected} to it.

A more accurate way to measure centrality is to consider \emph{network paths}.
A network path is a set of links that connects indirectly two nodes.
In the network representation of our rating environment a path of length $d$ is given by $d$ ordered agents who, pairwise, agreed on the same mark to assign to the same product.
In Figure \ref{fig:1}, considering all links, there is a path of length $3$ between $a2$ and $a3$ because agent $1$ picked $a2$ and agreed with agent $2$ on $b1$, then agent $2$ agreed with agent $5$ on $c2$, and finally agent $5$ picked $a3$.
The fact that this path passes through $b1$ and $c2$ assigns some structural \emph{centrality} to these two nodes.
Let us be more formal, the weight of the paths of length $1$ between any two nodes are represented by matrix $L$ itself, those of length $2$ are simply described by its square $L^2$, and so on, with longer paths that are exactly represented by higher powers of matrix $L$.
If we call $I$ the $aM \times aM$ identity matrix, and $\vec{1}$ the column vector made of $aM$ ones,
the Bonacich--Katz centrality $\vec{c}$ of a node is given by the implicit formula (see also \citealp{newman2004analysis} and \citealp{opsahl2010node})\footnote{%
It is also possible to assign centrality studying the spectral analysis of $L$ and assign weights with the eigenvectors corresponding to larger eigenvalues. However, as pointed out in \cite{bonacich2001eigenvector}, this is an approach that is much less robust to perturbations.
More on this in Section \ref{sec:beta}.}
\begin{eqnarray}
\vec{c} & = &  \beta L \vec{c}  \ \ , \nonumber   
\end{eqnarray}
that can be made explicit\footnote{%
As long as $\beta$ is not greater than the inverse of the maximum eigenvalue of $A$. Also this will be further discussed in Section \ref{sec:beta}.
}
by its unique solution
\begin{eqnarray}
\vec{c} & = &  \beta L \vec{1} + \beta^2 L^2 \vec{1} + \beta^3 L^3 \vec{1} + \dots \nonumber  \\
& = & \left( I - \beta L \right)^{-1} \vec{1} - \vec{1} \ \ . \label{eq:Bonacich}
\end{eqnarray}
Parameter $\beta>0$ plays the classical role of a multiplicator factor, and tells us how much we want to decrease the weight of longer paths.
Element $L \vec{1}$ in the second line of equation (\ref{eq:Bonacich}) is exactly the vector that counts the degree of each node, while the following elements consider larger paths.

\bigskip

So, to attribute a score to product $x$ one can make an average of all the possible marks for this product (i.e.~$x1$, $x2$, \dots, $xa$), weighted by their centrality:
\begin{equation}
\label{eq:explicit_centrality}
s_x = \frac{ \sum_{i=1}^a c_{xi} \cdot i }{\sum_{i=1}^a c_{xi}} \ \ .
\end{equation}
This score takes into account the aggregate information of the whole network and the correlations between the opinions in the overall poll of agents.
In this way we obtain a vector $\vec{s}$ that is our solution to the rating problem, and we call it the \emph{network centrality} (NC) method.
An important property of  equation (\ref{eq:explicit_centrality}) is that at the limit of  $\beta \rightarrow 0$ it coincides with the simple average, which is what would be obtained truncating the second line of equation (\ref{eq:Bonacich}) after the first element  $\beta L \vec{1}$.

\subsection{What is the best value for $\beta$?}
\label{sec:beta}

Variable $\beta$ represents the \emph{peer effect} between neighboring judgments in the adjacency matrix $L$ of all possible ranks for each product.
The best value for $\beta$ may clearly depend on the other variables of the problem, and in particular on the adjacency.
From the way it is built (equation (\ref{eq:link_value})), matrix $L$ is an $aM \times aM$ symmetric matrix with all non--negative entries.

It is well known (see e.g.~\citealp{BKD_AER}) that the strength of the peer effect depends on the largest eigenvalue of the adjacency matrix.
From the Perron-Frobenius theorem, the largest eigenvalue of $L$ is its unique positive eigenvalue $\lambda^+$, that lies in the interval $(0,N)$.\footnote{%
Actually, by the Perron--Frobenius theorem $\lambda^+$ is the unique strictly positive eigenvalue.
Consider now that the sum of the elements in each row of column $xi$ is the number of agents that actually rated good $x$ with score $i$, and this number is còearly bounded above by the total number $N$ of reviewers.
So, any eigenvalue of the matrix (which is how much the corresponding eigenvector is multiplied in the matrix product) cannot exceed $N$ in absolute value.}

One possibility is simply to balance the peer effect of the network structure with some $\beta$ that is inversely proportional to $N$, or to do it with a $\beta$ that is inversely proportional to the actual $\lambda^+$ of matrix $L$.
In the simulations of next section we try the following 6 values for $\beta$ (in decreasing order): $1 / \lambda^+$, $1/N$, $1/5N$, $1/10N$, $1/25N$ and $1/50N$.
Actually, when $\beta = 1 / \lambda^+$ equation (\ref{eq:Bonacich}) is not defined, because the matrix becomes singular.
However, at this limit $\vec{c}$ approximates  the eigenvector of $L$ corresponding to $\lambda^+$,\footnote{%
It is also well known, as discussed in \cite{bonacich2001eigenvector}, that this limiting result converges on a path that is extremely volatile to tiny fluctuations.}
 and this is what we compute in this case.

At the other side of the interval, at the limit $\beta \rightarrow 0$ the NC method will coincide with the average.
So, we conjecture that the larger the value of $\beta$, the better, but there is a trade off at $1 / \lambda^+$, which is the actual limit to stability of the infinite series in equation (\ref{eq:Bonacich}).
So, we try also an intermediate value between the first two, which is  $2/(\lambda^+ + N)$.

\subsection{Objective measures}
\label{sec:obj_measures}

How do we compare different rating methods?
Any method that aggregates the score from an $N \times M$ matrix of marks will result in an $M$ vector $\vec{s} \in [1,a]^M$, where $[1,a]$ is the set of real numbers in--between $1$ and $a$.  
When many data in the $N \times M$ matrix $\bf r$ of marks are missing, simple correlation between this vector $\vec{s}$ and the rows of the matrix are ambiguously defined and difficult to interpret.

\bigskip

One methos that has recently been proposed in the literature is the Separation--Deviation (SD) methos provided in equation (8a) of \cite{hochbaum2011rating}, which stems directly from the work of \cite{kemeny1962preference}.
We adapt it here to our notation.
They find the vector $\vec{s}$ that solves the following minimization problem
\begin{eqnarray}
\label{eq:hochbaum}
\min_{\vec{s}} & & \alpha \left( \sum_{m=1}^M \sum_{i=1}^{N-1} \sum_{j=i+1}^N w_{ij}^k \left( s_i - s_j - r_{i,m} + r_{j,m} \right)^2 \right) + 
 \left( \sum_{m=1}^M \sum_{i=1}^{N} v_i^k \left( s_i - r_{i,m} \right)^2 \right) \ \ ,  \\
\mbox{such that} & &  w_{ij}^k = \left\{
\begin{array}{rl}
1 & \mbox{if } r_{i,k} \ne - \mbox{ and } r_{j,k} \ne - \\
0 & \mbox{otherwise}
\end{array}
\right. 
\mbox{ \ and \ }
v_{i}^k = \left\{
\begin{array}{rl}
1 & \mbox{if } r_{i,k} \ne - \\
0 & \mbox{otherwise}
\end{array}
\right. \ \ ,
\nonumber 
\end{eqnarray}
where $\alpha$ is a positive real number that weights how much the first part of the objective function (the \emph{separation} penalty) is relatively  important with respect to the second part (the \emph{deviation} penalty).
We impose $\alpha=1$ and call \emph{SD measure} the objective function of the problem in (\ref{eq:hochbaum}).
This optimization problem is not trivial (more on this in Section \ref{sec:efficiency} below), but its solution is obtained from a system of linear equations, so it is generally unique.
However, this measure has a huge variance across different random realizations of matrix $\bf r$.
As we have checked in the simulations (see Section \ref{sec:simulations} below), the value of this objective function computed in the optimum and the value computed on the simple average are not different with statistical significance over random realizations of matrix $\bf r$.
Also, when we apply this measure to real data in Section \ref{sec:realdata} we observe that any method does not differ from any other with respect to this measure of more than $1\%$.

\bigskip

The measure that we will use to evaluate the performance of a measure $\vec{s}$ is the Kendall's Tau, as used in \cite{vanhoucke2010using}, where the relation between $\vec{s}$ and $\vec{r}_k$, on each couple of goods $i$ and $j$, is given by the following formula:
\begin{equation}
\label{eq:kendall_ij}
C_{ij} = \left\{
\begin{array}{rl}
0.5 & \mbox{if ($r_{k,i} = `-'$) or ($r_{k,j} = `-'$)} \\
& \mbox{otherwise} \\
1 & \mbox{if ($s_i< r_{k,i}$ and $s_j< r_{k,j}$) or ($s_i> r_{k,i}$ and $s_j> r_{k,j}$) or ($s_i= r_{k,i}$ and $s_j= r_{k,j}$)} \\
0.5 & \mbox{if ($s_i= r_{k,i}$ and $s_j \ne r_{k,j}$) or ($s_i \ne r_{k,i}$ and $s_j = r_{k,j}$)}  \\
0 & \mbox{otherwise}
\end{array}
\right. \ \ . \nonumber
\end{equation}
The Kendall Tau correlation between $\vec{s}$ and $\vec{r}_k$ is given by 
\begin{equation}
\label{eq:kendall_ij}
\tau_{k} = 4 \frac{\sum_{i=1}^{M-1} \sum_{j=i+1}^{M} C_{ij}}{M(M-1)} -1 ,
\end{equation}
which lies always between $-1$ and $1$.
And finally, the aggregate Kendall Tau correlation between $\vec{s}$ and $\bf r$ is given by the average  $\tau = \sum_{k=1}^N \tau_k / N$. \\
The analogy of this measure with the SD function of equation (\ref{eq:hochbaum}) are that absent marks have weight $0$ in the overall computation.
However, it differs in two ways: first, as the simple correlation, this measures increases with higher correlation; secondly, it is only an \emph{ordinal} measure, in the sense that only the ordering between numbers is important.
We show in next section, when presenting the output of our simulation exercise, that this measure has the necessary stability that guarantees identification of better methods.

\bigskip

We end this section with a simple example.

\begin{example}
\label{example:no_coherent}
Consider a simple case of three goods and four agents, depicted by the following table:
\begin{equation} \label{table:example}
\begin{array}{r|cccc|}
 & \multicolumn{4}{c}{Agents} \\
Products & 1 & 2 & 3 & 4 \\
\hline
a & 4 & - & 1 & 1  \\
b & - & 2 & 2 & 3 \\
c & 5 & 1 & - & 2 \\
\hline
\end{array} \ \ .
\end{equation}

\bigskip

In this case, averaging, we assign values $(2,2.33,2.67)$ to the goods, and this conflicts with the rankings of agents $2$ and $4$.
A ranking that instead value first good $2$, then good $3$ and finally good $1$ would respect the order of each agent.
If we look at the network representation of Figure \ref{fig:example2}, analogous to the one in Figure \ref{fig:1}, we see that agent 1 is just an isolated component of the network, so that its scores have little in common with other agents. 
\begin{figure}[h]
\begin{center}
 \includegraphics[width=10cm]{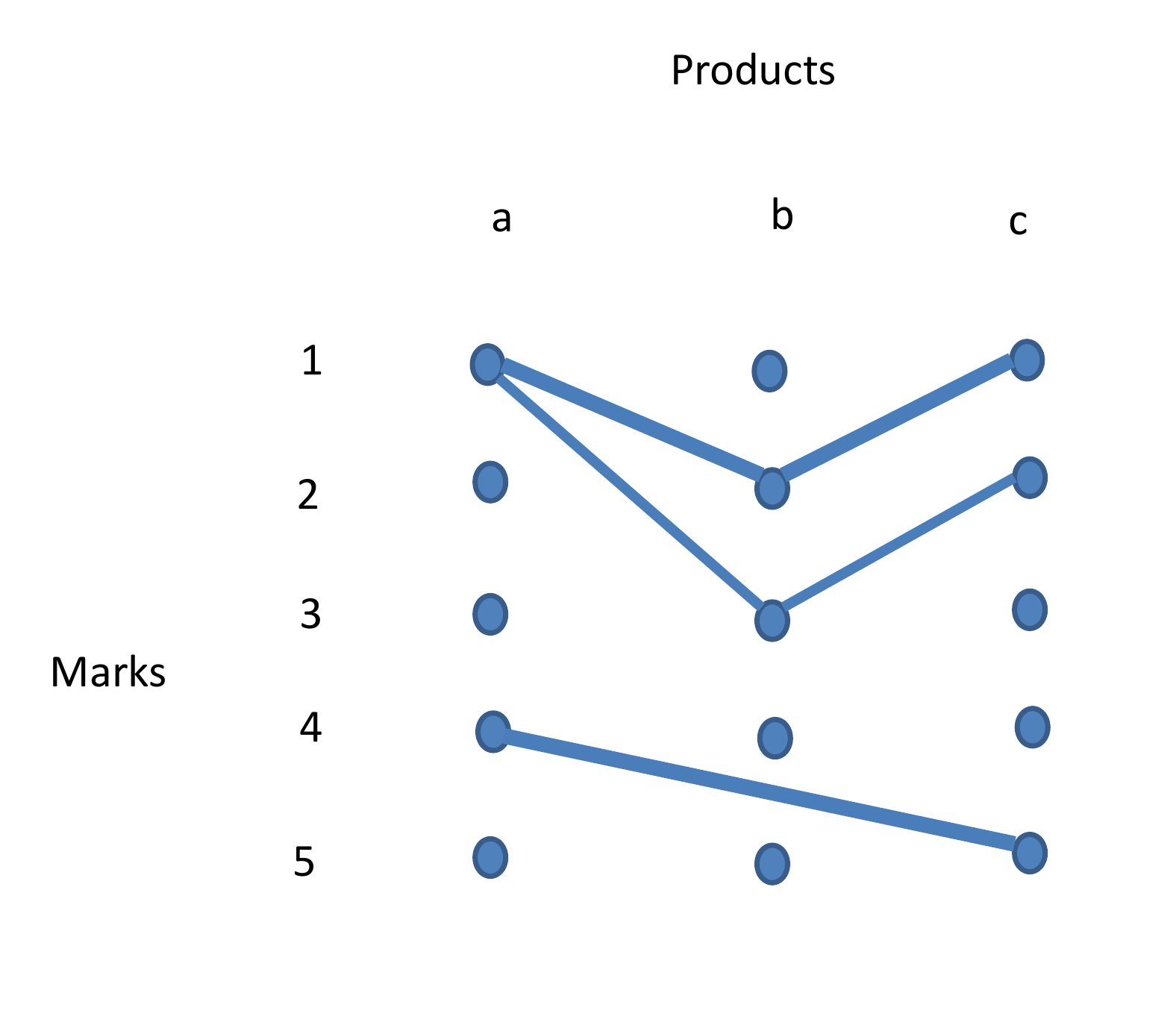}
\caption{The network representation of Table \ref{table:example} (bolder lines have higher weight).} \label{fig:example2}
\end{center}
\end{figure}

Table \ref{table:example_measures} shows the outcome of the SD method and of  the simple average, with respect to our NC method, with the values of $\beta$ listed in Section \ref{sec:beta} (in this example $\lambda^+$ is $1.560$ and $N=4$).
It is clear from here that as $\beta \rightarrow 0$ we asymptotically approximate the average.
The last two lines report the SD measure and the Kendall's Tau of each method.
The NC method with $\beta = 2/(\lambda^+ +N)$ is the second best with respect to the SD measure (which is the measure that the SD method minimizes by definition).
The two NC methods with higher beta's are also those that preserve the correct ordering in the ranking of the products, as shown by the Kendall's Tau.
This example anticipates the results that we obtain from simulations and from an application to real data in Sections \ref{sec:simulations} and \ref{sec:realdata}.

\setcounter{table}{8}
\begin{table}[htbp]
  \centering
  \caption{Different methods applied to the example of Table \ref{table:example}.}
    \begin{tabular}{c|cc|ccccccc}
    \hline
    products & SD    & Avg   & $\beta=1/\lambda^+$  & $2/(\lambda^+ +N)$ & $1/N$ & $1/5N$ & $1/10N$ & $1/25N$ & $1/50N$ \\
    \hline
    a     & 1.667 & 2.000 & 1.000 & 1.783 & 1.879 & 1.982 & 1.991 & 1.997 & 1.998 \\
    b     & 2.889 & 2.333 & 2.302 & 2.325 & 2.330 & 2.333 & 2.333 & 2.333 & 2.333 \\
    c     & 2.444 & 2.667 & 1.403 & 2.322 & 2.460 & 2.632 & 2.650 & 2.660 & 2.663 \\
\hline
    SD measure & 18.111 & 22.380 & 24.438 & 20.580 & 21.111 & 22.100 & 22.218 & 22.292 & 22.310 \\
    $\tau_{Kendall}$   & 0.5   & 0.1667 & 0.5   & 0.5   & 0.1667 & 0.1667 & 0.1667 & 0.1667 & 0.1667 \\
    \hline
    \end{tabular}%
\label{table:example_measures}
\end{table}%

\end{example}

\clearpage

\subsection{Computational efficiency}
\label{sec:efficiency}

The infinite sum in the second line of equation (\ref{eq:Bonacich}) can be truncated at the $i^{th}$ step, and so its computational cost can be arbitrarily reduced at the expenses of accuracy.
In fact, at the limit $\beta \rightarrow 0$ we have the truncation $i=1$ and the NC method coincides with the average, whose computational cost is linear. 
However, the infinite series is perfectly computed in the third line, and matrix inversion is a very well studied problem.
Actually, \cite{williams2012multiplying} has recently proposed an algorithm that computes the inverse of a general $n \times n$ matrix in $O \left( n^{2.373} \right)$ computational time.\footnote{%
Even if faster algorithms provide \emph{good enough} approximate solutions when the original matrix is \emph{sufficiently} sparse. On this see e.g.~\cite{amestoy2012computing}.}
So, our method needs to invert an $aM \times aM$ matrix, where $a$ is a constant, and can then be solved in $O \left( M^{2.373} \right)$ computational time.

\bigskip

\cite{hochbaum2006methodologies} and \cite{hochbaum2011rating} discuss how their SD method, from equation (\ref{eq:hochbaum}), can be solved in $O(M N \log(N^2/M) \log N)$ time, constructing first an $NM \times NM$ adjacency matrix, and then applying the \emph{minimum cut} problem to the network resulting from that matrix,  with the algorithm proposed by \cite{ahuja2003solving}.
When $N$ is large, and even exceeds $M^{1.373}$ (as can easily be the case in the online applications that we have in mind) their method is clearly slower and requires much more memory than the one we propose.

\section{Simulations}
\label{sec:simulations}

We test the quality of our method, with the values of $\beta$ listed in Section \ref{sec:beta}, with respect to the SD method and to the simple average, on synthetic data generated in the following way.
We consider $a=5$ (so, five possible scores) and $N=10$ (ten agents).
For $M$, the number of goods, we have two cases: $M=10$ and $M=50$.
Each consumer $i$ rates randomly, with $i.i.d.$ probabilities, in the following way:
each good $j$ is not rated with probability $1-p$ (so that $r_{i,j}=`-'$) and rated with probability $p$.
When rated, $r_{i,j}$ has on of the 5 possible values from $\{ 1,2,3,4,5 \}$ with uniform probabilities.
In  this way every single element of the matrix $r_{i,j}$ is i.i.d. wit respect of all the others, and a fraction $1-p$ of them are expected to be empty: $`-'$.
First, we analyze how the different rating  methods perform with respect to the Kendall's Tau correlation from equation (\ref{eq:kendall_ij}), in the two cases with $N=10$ and $N=50$.
Then, in the case $M=50$, we check for robustness of different measures when additional agents with systematically biased reports are added to the sample.

\subsection{Kendall's tau correlation}
\label{sec:sim_Kendall}

For both cases $M=10$ and $M=50$ we generate $200$ i.i.d. realizations of $\bf r$, for 13 evenly spaced values of $p$ in--between $0.4$ (less than half of the goods are expected to be reported by each agent) and $1$ (all goods are reported by each agent).
For each realization we compute the average mark for each good, the measure resulting from the SD method of \cite{hochbaum2011rating}, and our centrality measure with respect to the 7 different values of $\beta$ discussed in Section \ref{sec:beta}: largest eigenvalue $\lambda^+$ of the $L$ matrix, $1/N$, $2/(\lambda^+ + N)$, $1/5N$, $1/10N$, $1/25N$ and $1/50N$.
Finally, for each of these $M$--dimensional vectors of measures, we compute the Kendall's Tau correlation from equation (\ref{eq:kendall_ij}).\footnote{%
The Matlab codes are available at \url{http://www.econ-pol.unisi.it/paolopin/WP/codes_LPW14.txt}.}

\begin{figure}[h]
\begin{center}
 \includegraphics[width=16cm]{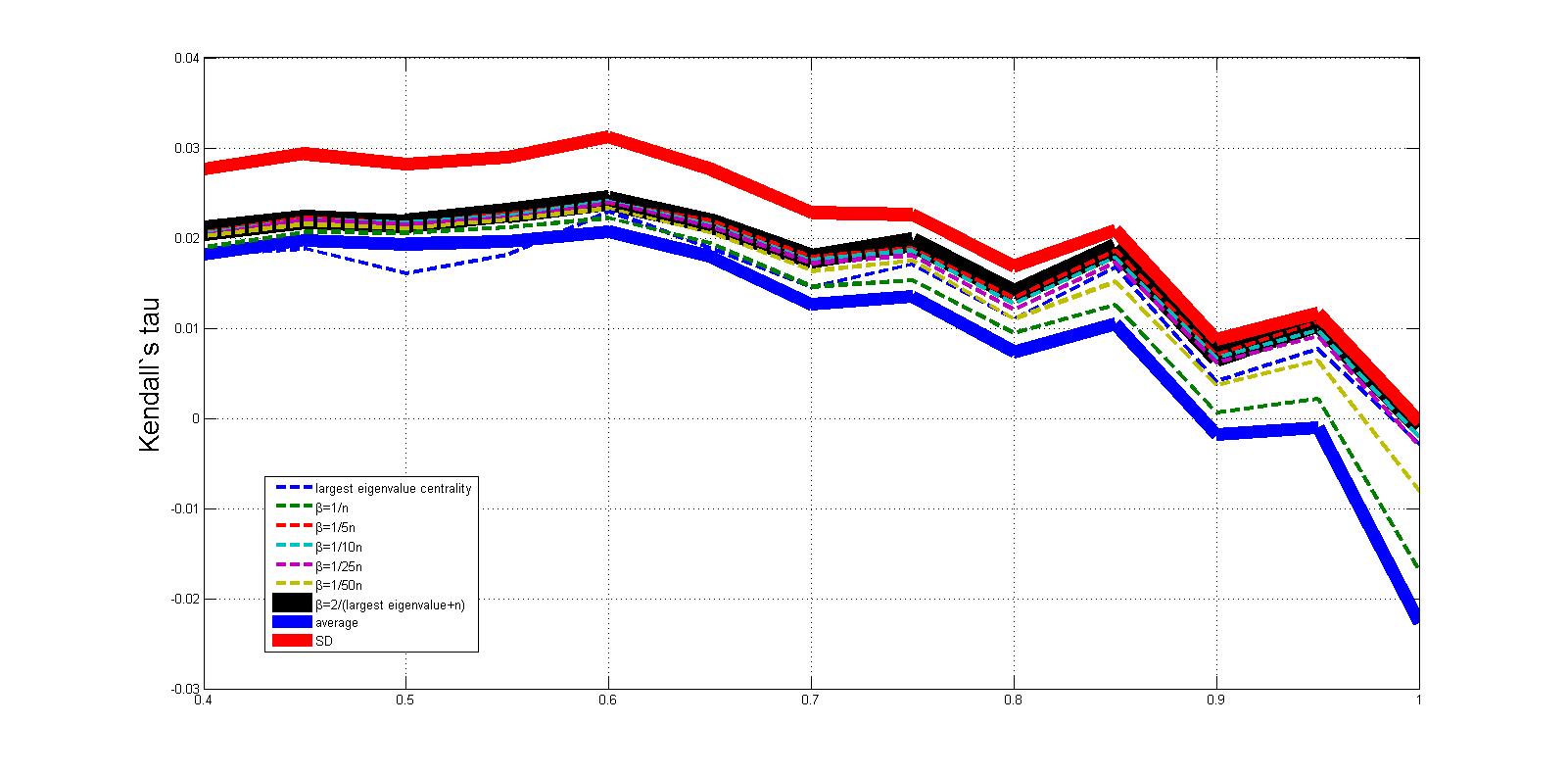}
 \includegraphics[width=16cm]{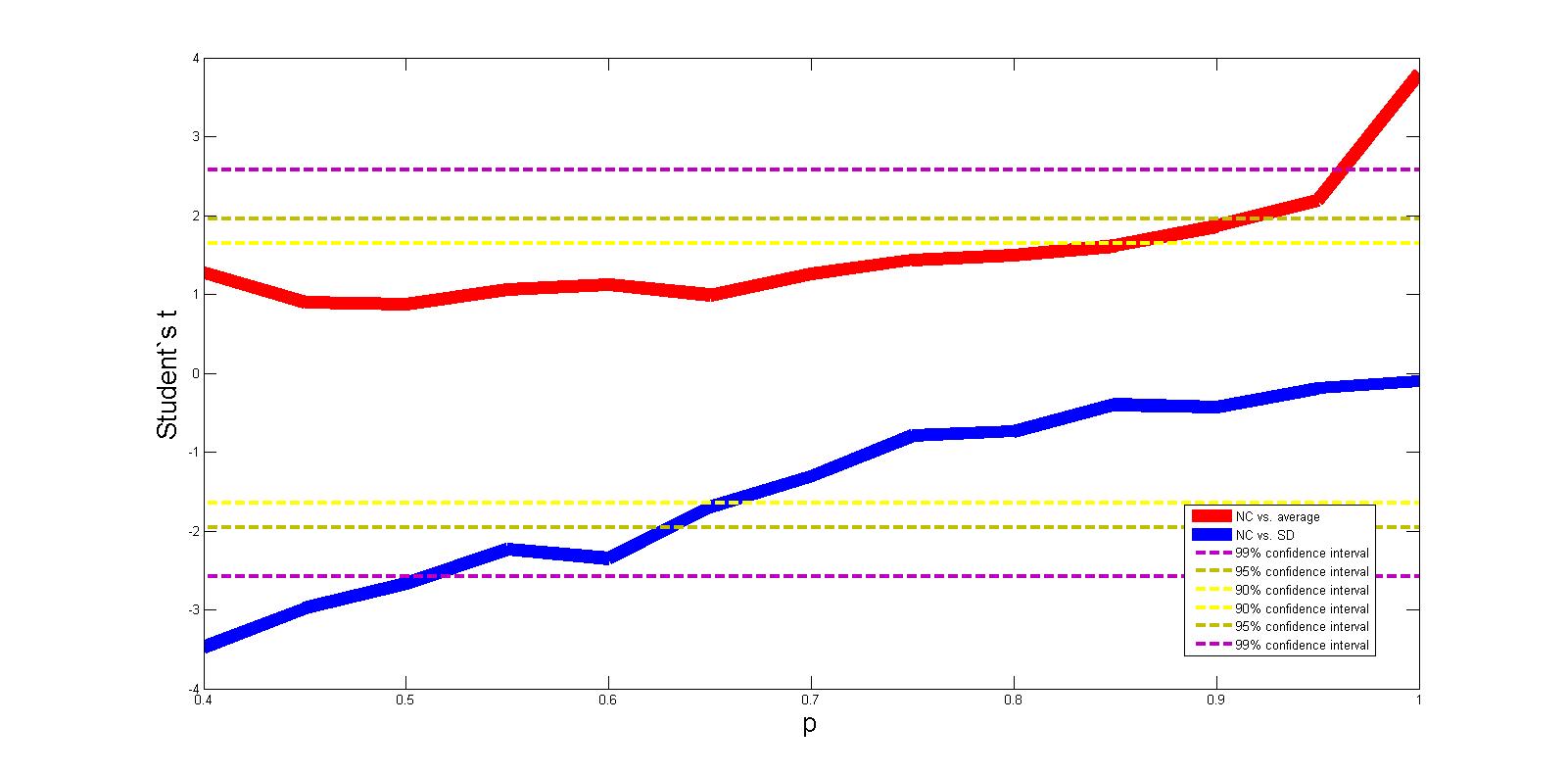}
\caption{Average outcome of the Kendall's tau measure on the simulations, with 9 different methods, $p$ varying from $0.4$ to $1$, and $M=10$.
Lower plot shows Student's $t$ comparison between one of the NC methods with respect to average and SD.} \label{m10plot}
\end{center}
\end{figure}

\begin{figure}[h]
\begin{center}
 \includegraphics[width=16cm]{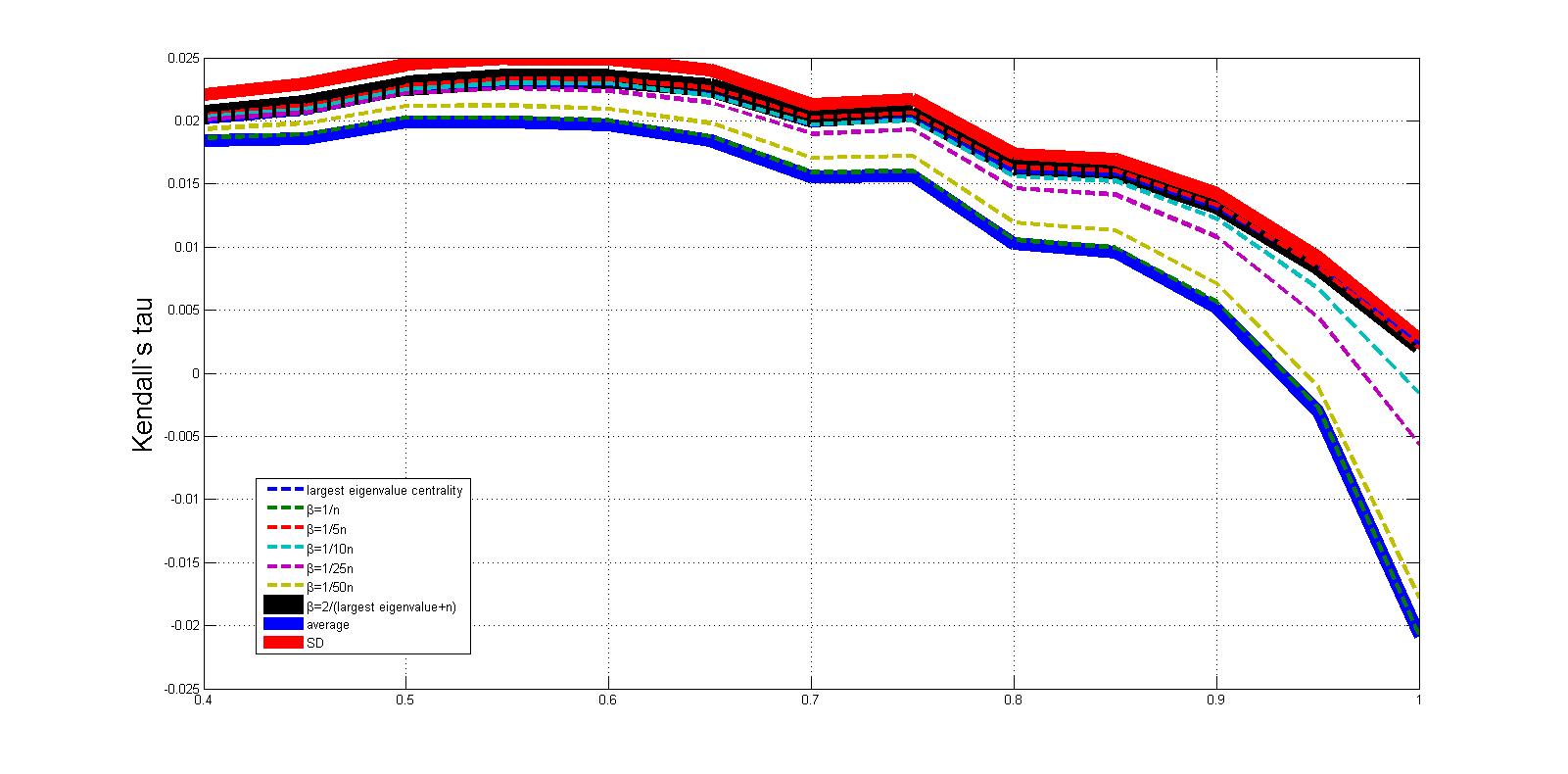}
 \includegraphics[width=16cm]{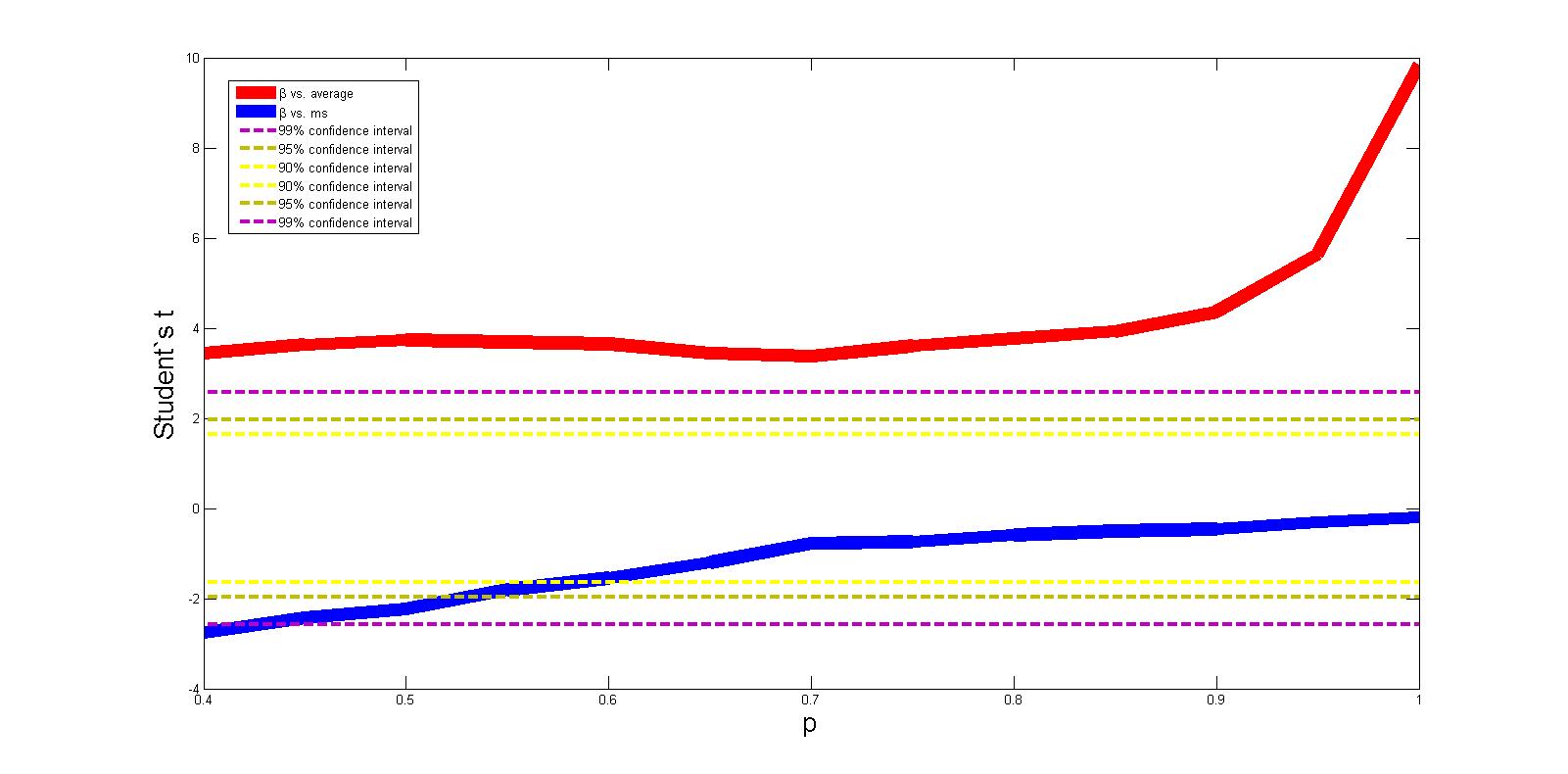}
\caption{Average outcome of the Kendall's tau measure on the simulations, with 9 different methods, $p$ varying from $0.4$ to $1$, and $M=50$.
Lower plot shows Student's $t$ comparison between one of the NC methods with respect to average and SD.}  \label{m50plot}
\end{center}
\end{figure}

Results of the average outcomes are reported in the upper parts of Figures \ref{m10plot} (for $M=10$) and \ref{m50plot} (for $M=50$).
From these average trends, it comes out clearly out that SD is the best performing measure, and that average is the worse, while the centrality measure with different values of $\beta$ lie in--between.
The best value of $\beta$ seems to be $2/(\lambda^+ + N)$.
However, we need to take variance into account when analyzing these results.
In \ref{app:boxplots}, Figures \ref{m10boxplot}  (for $M=10$) and \ref{m50boxplot} (for $M=50$) report the boxplots of all the $200$ realizations for the following three measures: DS, average, and NC measure with $\beta=2/(\lambda^+ + N)$.
The lower parts of Figures \ref{m10plot} (for $M=10$) and \ref{m50plot} (for $M=50$) take also variance into account and plot the Student's $t$--test to check if the centrality measure with $\beta=2/(\lambda^+ + N)$ is statistically different from the SD measure and the simple average, as $p$ changes.
When $M=10$ the NC measure is not statistically different from the other two measures for most values of $p$: it performs significantly better than the average for high $p$, and significantly worse than the MS measure for low $p$.
But when $M=50$ the NC method is always better than the average with $99\%$ statistical confidence, while it is not statistically different form the MS measure for $p$ above $0.6$.

\bigskip

We have tried, on the same set of simulations, also two other measures of coherence: the objective measure from equation (\ref{eq:hochbaum}) and the simple average correlation of a method with those of all the agents (limited for each agent to those goods that are actually rated by that agent).
However, those two measures have a much larger variance than the Kendall's tau correlation, and all the outputs are not statistically different, according to the Student's $t$--test, in any of the points presented in Figures  \ref{m10boxplot} and \ref{m50boxplot}.
The reason is clearly that while Kendall's tau correlation is penalized only by violations in the ordering, and is hence \emph{ordinal}.
the measure from equation (\ref{eq:hochbaum}) and the simple correlation are cardinal measures that are affected also by the magnitude of the scores.

\clearpage

\subsection{Robustness to biased fake data}
\label{sec:sim_consistency}

Now we pose a different question: what happens to the ranking measure that we are using if we add fictitious agents that adopt a systematically biased report?
To do so we consider the previous case with $N=10$ and $M=50$, and we add $H$ agents (with $H$ from 1 to 5) to the original 10 ones.
These agents just assign mark 1 to the first three goods in the $M$ list, and mark 5 to the last three ones (it is clear that the order of the goods plays no role, and the point is just that some goods are systematically rated at the top, while others are systematically rated at the bottom).

In this scenario it is not clear which measure preserves better the original ranking from the inclusion of fake reviewers.
In principle, a measure should detect that those new agents are somehow \emph{different} from the original ones.
The NC measure does exactly this, because the sub--network generated by the new fictitious agents will be an almost disconnected part with respect to the original network.

\begin{figure}[h]
\begin{center}
 \includegraphics[width=16cm]{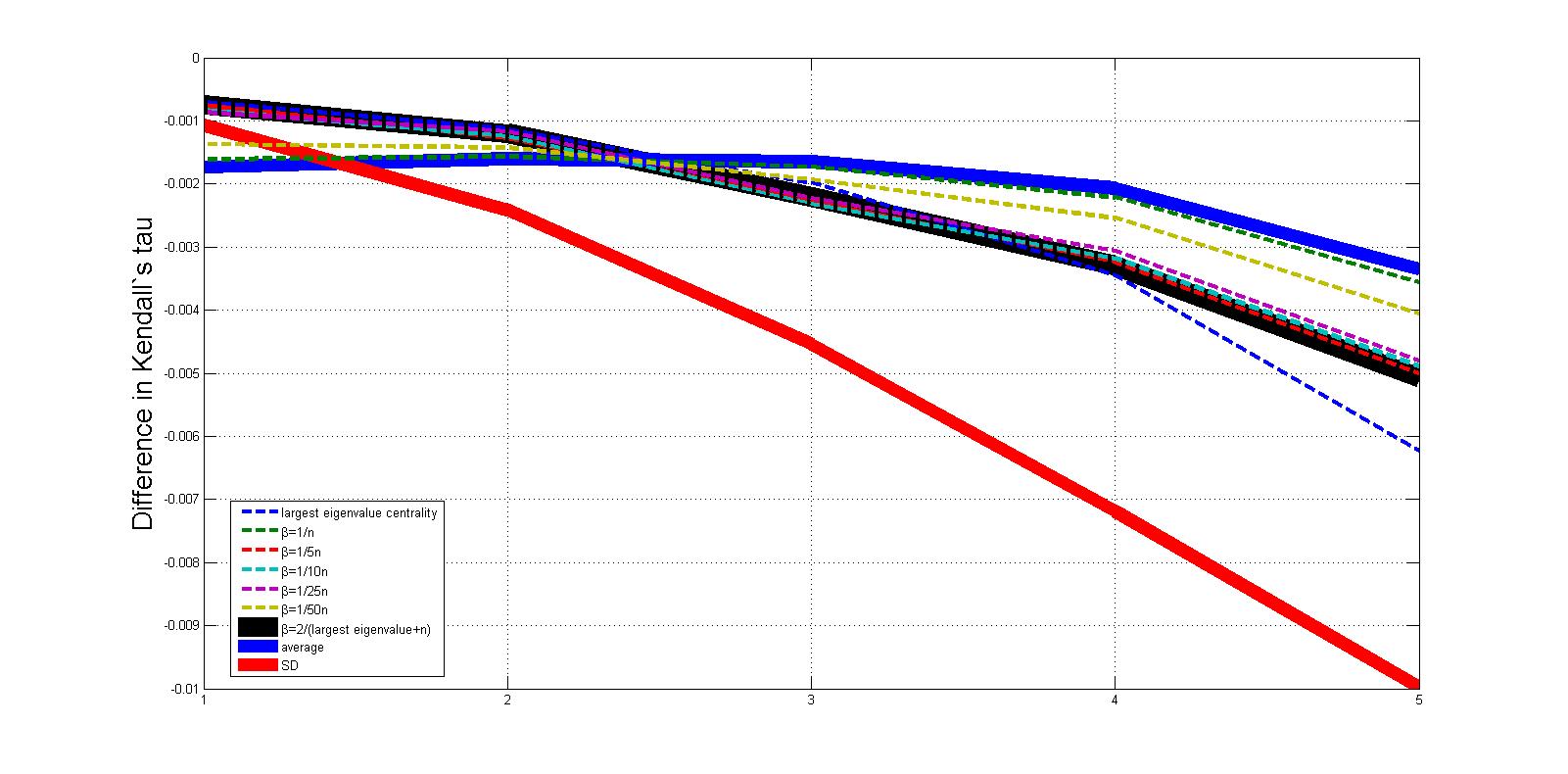}
 \includegraphics[width=16cm]{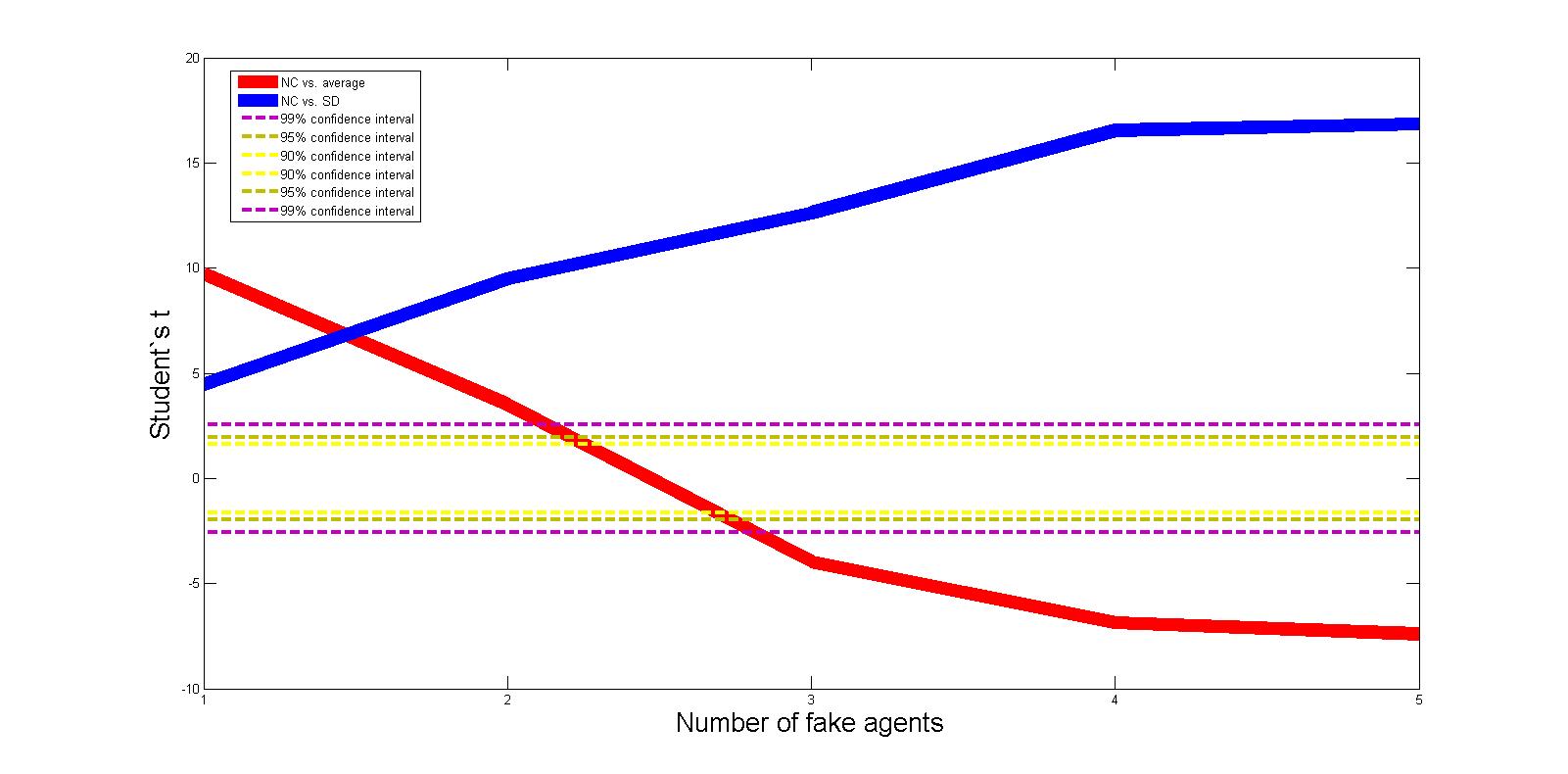}
\caption{Average outcome in the simulations of the difference in Kendall's tau measure, between the original data and those obtained adding $X$ fake.
There are 9 different methods, and $X$ varies from $1$ to $5$.
Lower plot shows Student's $t$ comparison between one of the NC methods with respect to average and SD.}\label{fake_plot}
\end{center}
\end{figure}

As a measure for our simulations we use the difference between the Kendall's tau correlation index computed on the measure obtained aggregating all the consumers, and the Kendall's tau correlation index computed on the measure that would have been obtained only from the original reviewers.
Clearly, this difference will be negative, but a good measure will be one that minimizes it in absolute value.
With the same notation of last section, Figure \ref{fake_plot} reports the outcome and the $t$--test (and Figure \ref{fake_boxplot} in \ref{app:boxplots} the boxplots).

The results speaks in great favor of our NC method: when up to 2 fake reviewers are added to the original 10 ones, the NC measure with $\beta=2/(\lambda^+ + N)$ is better than any other measure with high statistical significance.
When more fake agents are added it improves its outcome with respect to the SD measure, but it becomes worse than the average with more than $99\%$ statistical significance.

\clearpage

\section{Real data}
\label{sec:realdata}

Finally, we test our method, with the values of $\beta$ listed in Section \ref{sec:beta}, with respect to the SD method and to the simple average, on a real dataset.
In this case we consider all measures of efficiency, and as it is a single realization we will not be able to compute statistical significance when the outcomes are different.

We use a dataset recording rating information on movies provided by the site \url{grouplens.org}, from the University of Minnesota.\footnote{%
The actual dataset is called \emph{movielens} and is part of an academic project of the University of Minnesota on \emph{Social Computing}.
`Movielens' is presented as a tool for rating and comparing movies: anyone can register, rate and consult all the movies from a fixed list.
The version we analyze has been downloaded by us in December 2013, and are available at \url{http://www.econ-pol.unisi.it/paolopin/WP/movielens_LPW14.data}.} 
The data set encompasses 943 customers and 1682 movies. It records the rating scores of each movie rated by its customers. All the movies are rated from score 1 to score 5, with 1 as the worst and 5 as the best. The total amount of rating records are about 100,000, so each customer rated on average about 106 movies which is far less than 1,682. Such sparse data would pose a challenge for many of the existing evaluation methods. 
Also, the SD method would need to construct a matrix of size $(NM)^2>2.5\cdot 10^{12}$, in order to apply the minimum cut algorithm of \cite{ahuja2003solving}.\footnote{%
We have computed directly with \emph{Matlab} the direct optimization of the objective function in (\ref{eq:hochbaum}).}
For our NC method we need instead to invert a matrix of size $(5M)^2 \simeq 7 \cdot 10^7$, which is tractable.

\begin{table}[htbp]
  \centering
  \caption{Different methods applied to a real dataset.}
    \begin{tabular}{rccc}
    \toprule
          &      `Movielens' dataset      &  \\
    \midrule
    Method & $\tau_{Kendall} \cdot 10^{-5}$ & Correlation   &  SD value $\cdot 10^7$ \\
    $\beta=1/\lambda^+$ & 0.48  & 0.4256 & 1.699964 \\
     $\beta=2/(\lambda^+ +N)$ & 0.61  & 0.4208 & 1.692521 \\
     $\beta=1/5N$ & 0.55  & 0.4205 & 1.692400  \\
     $\beta=1/10N$ & 0.53  & 0.4202 & 1.692308 \\
     $\beta=1/25N$ & 0.52  & 0.4202 & 1.692302 \\
     $\beta=1/50N$ & 0.52  & 0.4202 & 1.692294 \\
\hline
    Average & 0.44  & 0.4202 & 1.692244 \\
    SD    & 0.63  & 0.4099 & 1.691011 \\
    \bottomrule
    \end{tabular}%
  \label{tab:addlabel}%
\end{table}%

The results are consistent with those of the simulations.
The NC method with $\beta=2/(\lambda^+ +N)$ performs almost as well as the SD method with respect to  the Kendall's tau correlation.
It is surprising that the SD method is actually the worst one with respect to the simple correlation, but we have discussed in Section \ref{sec:obj_measures} how it is difficult to interpret it when many missing data are present, as in this case.
Finally, the SD method, by definition minimizes the objective function from equation (\ref{eq:hochbaum}), and with this respect the NC method is not better than the simple average, even if all the numbers are very similar and there is no clear added value in not adopting the simple average.
Actually,  in relative terms the difference between the NC method with $\beta=1/\lambda^+$ and the SD method is about $0.5\%$.

\section{Conclusion}
\label{sec:conclusion}

We have provided a network centrality rating method for aggregating the overall information of consumers rating products.
We argue that it is an optimal trade--off between computational efficiency and desirable features, especially when compared with the simple average and with other methods proposed in the literature.
However, the methods actually used by online sites are not evident, and consumers can only perceive them as black--box.
The algorithms implemented there are probably very sophisticated, and make use of many more information than those we have considered in the present paper.
For example, the popular site \url{www.tripadvisor.com}, that is actually based only on its rating service, declare that it gives different weights to different reports depending on the \emph{importance} of the reviewer and on the timing, attributing higher value to more recent reports.\footnote{%
This is for example explicitely stated in the following two urls: \url{http://help.tripadvisor.com/articles/200613987} and \url{http://help.tripadvisor.com/articles/200614027}.}
The service of aggregating ratings could be also customized upon request, as the \emph{importance} could be assigned in a way that reviewers with certain characteristics are given more or less weights depending on who performs the request.
What we want to stress here is that attributing different weights to consumers or to single reports is something that can easily be done in any of the methods we have compared, including the average and our NC method.
Actually, showing that at the limit of $\beta \rightarrow 0$ our NC method actually coincide with the average, we provide continuity between any weighted average method and the corresponding NC method with $\beta>0$.
So, the \emph{horse--race} exercise that we have performed would not be affected by this extra differentiation.

Also, we have shown that our NC method is, in some cases, the best one in neglecting the score of systematically biased fake reports.
It is clear that knowing who this fake and/or biased consumers are will help in disregarding their value, as one could give them less or even zero weight, but our NC method does it generically by its own nature, starting ex--ante with equal weights and no extra information.

So, any other additional information or algorithm that can distinguish the importance of reviewers and even single reports will be helpful for our method, as for any other one, but it is orthogonal to the properties that we have shown in the present paper.
In this sense, we provide an additional tool that can be implemented in real world applications, and we show that its properties are very useful in obtaining an aggregate rating measure that preserves the original ranking made by reviewers, and is able to detect, without any additional information, those reviewers that are under suspicion of being not genuine.

\bibliographystyle{chicago}
\bibliography{rating}


\setcounter{section}{1}

\appendix
\global\long\def\thesection{Appendix \Alph{section}}
 \global\long\def\thesubsection{\Alph{section}.\arabic{subsection}}
 \setcounter{equation}{0} \global\long\def\theequation{\alph{equation}}
 \setcounter{proposition}{0} \global\long\def\theproposition{\Alph{proposition}}
 \setcounter{definition}{0} \global\long\def\thedefinition{\Alph{definition}}

\section{Boxplots from the simulations}
\label{app:boxplots}

For the three sets of simulations discussed in Section \ref{sec:simulations}, we report the boxplots of all the 200 outcomes on each set of variables, for the NC method with $\beta=2/(\lambda^+ +N)$, the SD method, and the simple average.
On the $y$--axis we report all the cases, and on the $x$--axis we report the measures: Kendall's tau for the simulations discussed in Section \ref{sec:sim_Kendall} (Figures \ref{m10boxplot} and \ref{m50boxplot}) and the difference between Kendall's tau measures, that we call \emph{robustness}, for the simulation discussed in Section \ref{sec:sim_consistency} (Figure \ref{fake_boxplot}).
It is clear that there is a large variance and many realizations overlap, and this is why we show in the figures of the main text also the outcome of Student's $t$--tests to check when these outcomes are statistically different.

\begin{landscape}

\begin{figure}[h]
\begin{center}
 \includegraphics[width=22.6cm]{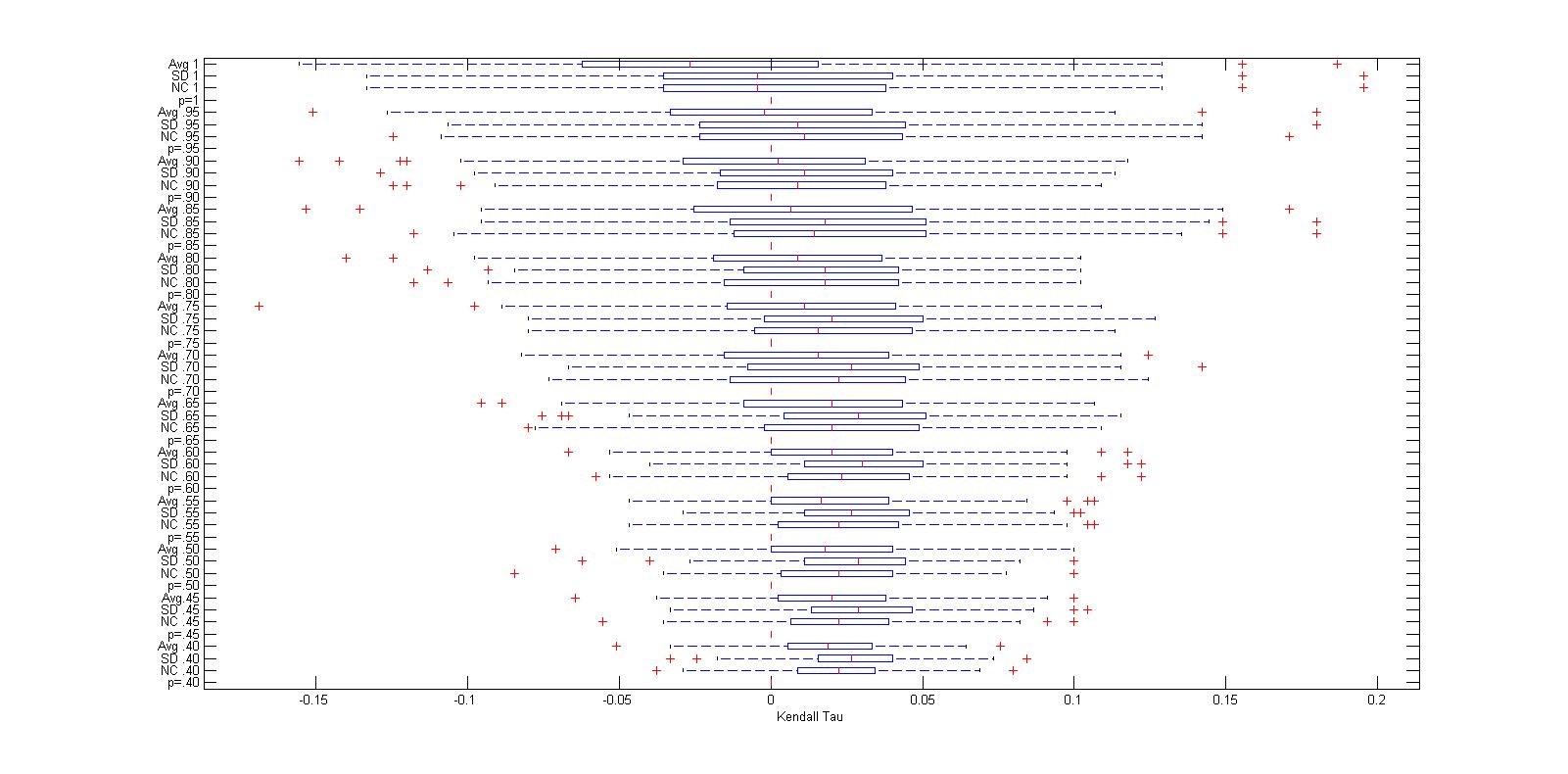}
\caption{Full boxplot of the simulation summarized in Figure \ref{m10plot}, for the average, the SD method, and the NC methoid with $\beta=2/(\lambda^+ +N)$.} \label{m10boxplot}
\end{center}
\end{figure}

\begin{figure}[h]
\begin{center}
 \includegraphics[width=22.6cm]{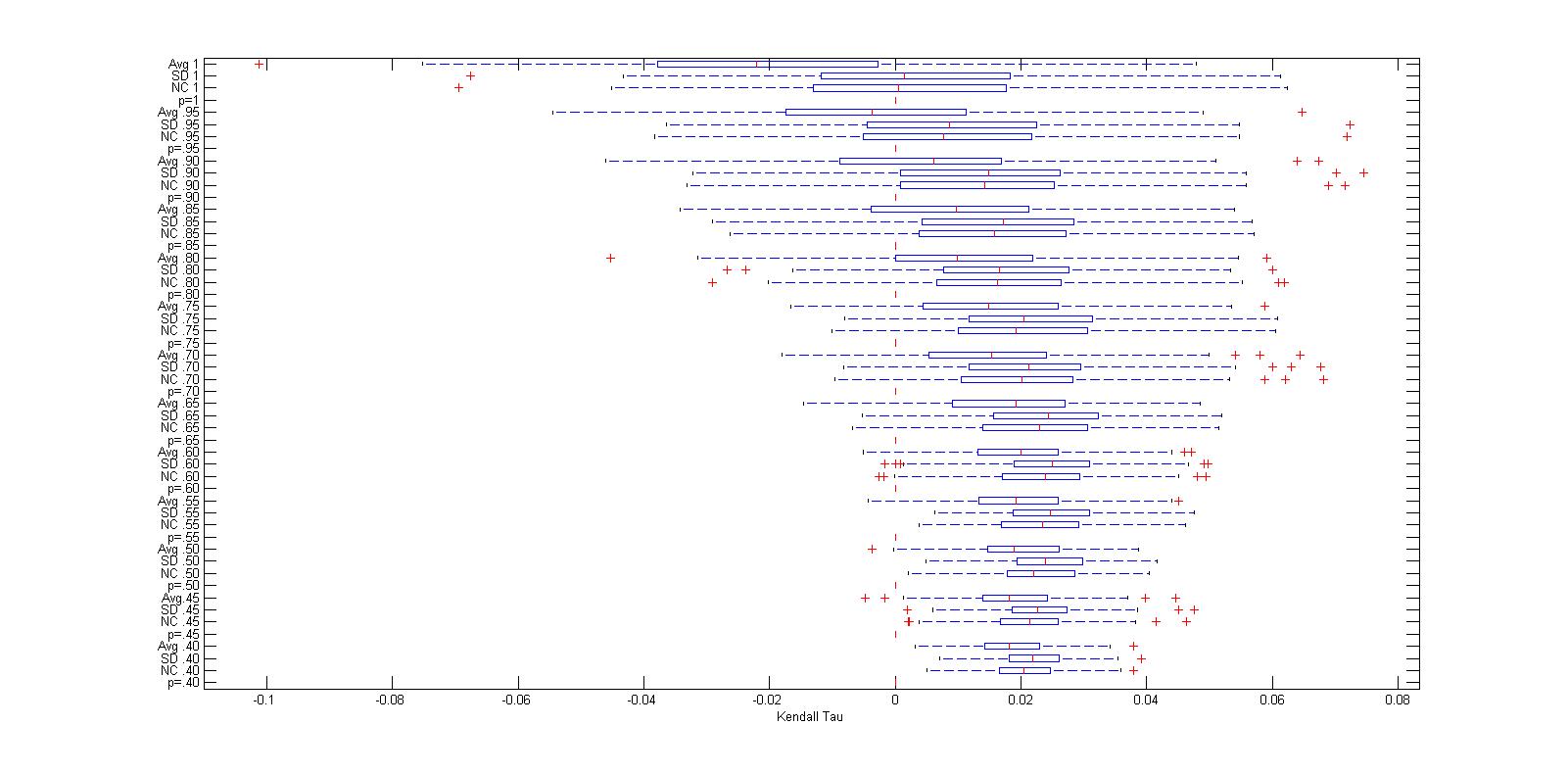}
\caption{Full boxplot of the simulation summarized in Figure \ref{m50plot}, for the average, the SD method, and the NC methoid with $\beta=2/(\lambda^+ +N)$.} \label{m50boxplot}
\end{center}
\end{figure}

\clearpage

%

\begin{figure}[h]
\begin{center}
 \includegraphics[width=22.6cm]{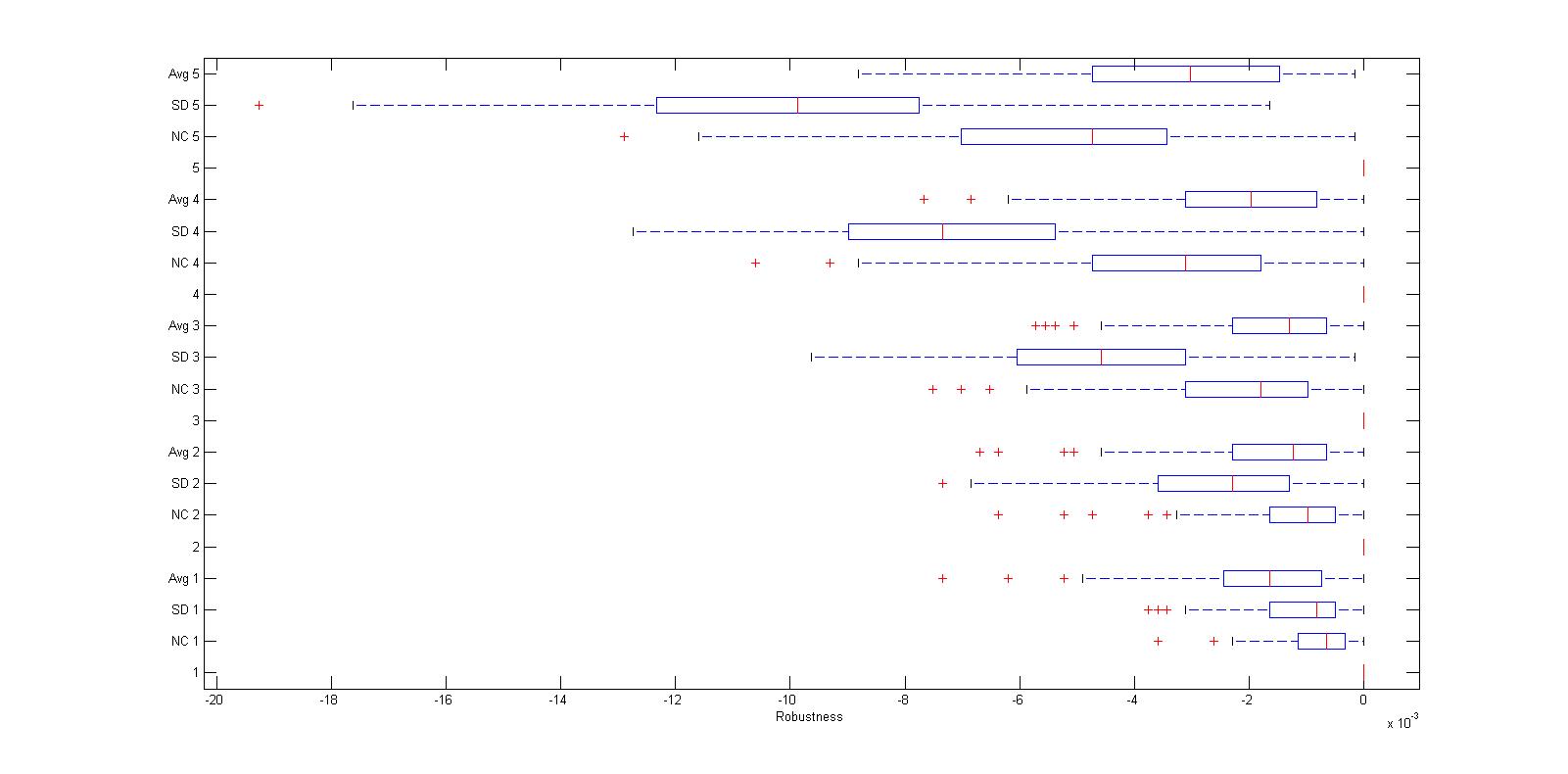}
\caption{Full boxplot of the simulation summarized in Figure \ref{fake_plot}, for the average, the SD method, and the NC methoid with $\beta=2/(\lambda^+ +N)$.} \label{fake_boxplot}
\end{center}
\end{figure}

\end{landscape}

\enddocument